\documentclass[12pt]{article}

\newcommand{\be}{\begin{equation}}
\newcommand{\ee}{\end{equation}}
\newcommand{\bn}{\begin{eqnarray}}
\newcommand{\en}{\end{eqnarray}}

\def\vE{\vec{E}}
\def\VE*{\vec{E}^{*}}

\def\vA{\vec{A}}
\def\vx{\vec{x}}

\def\nn{\nonumber}
\def\ba{\begin{eqnarray}}
\def\ea{\end{eqnarray}}
\def\ni{\noindent}
\begin{document}
\begin{center}
\noindent{\large \textbf{{\Large Dirac-like Monopoles in Three Dimensions
and Their\vspace{3mm} Possible Influences on the Dynamics of Particles}}} 
\vspace{3mm}

\noindent{\ 
{\Large E. M. C. Abreu$^{a}$, J. A. Helay$\ddot{\mbox{\rm 
e}}$l-Neto$^{b}$, $^{c}$, M. Hott$^{a}$ \\[0pt]and W. A. 
Moura-Melo$^{d}$} 
} \vspace{3mm}

\noindent 

{\large $^a$ \textit{Departamento de F\'{\i}sica e Qu\'{\i}mica,
UNESP/Guaratinguet\'a, \\[0pt]
PO Box: 205, Guaratinguet\'a, 12516-410, SP, Brazil,} \\[0pt]
\textsf{E-mail: everton@feg.unesp.br and hott@feg.unesp.br.}\\[0pt]
$^b$ \textit{Centro Brasileiro de Pesquisas F\'{\i}sicas \\[0pt]Rua Xavier
Sigaud 150, Urca, 22290-180, Rio de Janeiro, RJ, Brazil,} \\[0pt]
$^c$ \textit{Grupo de F\'{\i}sica Te\'orica, Universidade Cat\'olica de
Petr\'opolis\\[0pt]Rua Bar\~ao do Amazonas 124, 25685-070, 
Petr\'opolis, RJ, Brazil,} \\[0pt] \textsf{E-mail: helayel@cbpf.br and 
helayel@gft.ucp.br} \\[0pt] $^d$ \textit{Departamento de Ci\^encias 
Exatas, Universidade Federal de Lavras, \\[0pt] Caixa Postal 37, 
37200-000, Lavras, MG, Brazil,} \\[0pt] \textsf{E-mail: 
winder@stout.ufla.br.}\\[0pt] } 
\vspace{1mm}

\today
\end{center}

\begin{abstract}
\ni Dirac-like monopoles are studied in three-dimensional Abelian Maxwell
and Maxwell-Chern-Simons models. Their scalar nature is highlighted 
and discussed through a dimensional reduction of four-dimensional
electrodynamics with electric and magnetic sources. Some general properties
and similarities of them whenever considered in Minkowski or Euclidian space
are mentioned. However, by virtue of the structure
of the space-time in which they are studied a number of differences 
among them take place. Furthermore, we pay attention to some
consequences of these objects whenever acting upon usual particles. Among other
subjects, special attention is given to the study of a Lorentz-violating
non-minimal coupling between neutral fermions and the field generated by a
monopole alone. In addition, an analogue of the Aharonov-Casher effect
is discussed in this framework.
\end{abstract}

\section{Introduction and Motivation}

The idea that magnetic monopoles, as stable particles carrying 
magnetic charges, ought to exist has proved to be remarkably durable. 
In (3+1) dimensions, a persuasive argument was first put forward by 
Dirac in 1931 \cite{dirac}, who invoked such objects in order to 
provide a theoretical explanation why electric charges appear only as 
multiples of the elementary one.\newline

Furthermore, 't Hooft \cite{thooft} and Polyakov \cite{polyakov} 
discovered that the existence of magnetic monopoles follows from quite 
general ideas about the unification of the fundamental interactions. 
Nowadays, it is well-known that such objects emerge from general 
``grand unified'' theories of particle physics 
whose gauge group is suitably broken-down to the $U(1)$-factor. Indeed, Dirac 
had proved the consistency of structureless magnetic monopoles with quantum 
electrodynamics. On the other hand, some properties of the 't Hooft-Polyakov 
monopole, such as its size and mass, are determined by the distance scale of the 
spontaneous symmetry breakdown of a grand unified theory. The magnetic charge, 
$g$, of the monopole is typically the ``Dirac charge'', $g_{D}=1/2e$, 
which is distributed over a core with a radius of order $ M_{X}^{-1} 
$ (the unification distance scale) while its mass is 
comparable to the magnetostatic potential energy of the core. An 
excellent review on these subjects may be found in Ref.\cite{go}.\\

In turn, the study of 
three-dimensional field theories has attracted a great deal of
efforts since nearly two decades \cite{jt,djt}. Even though such studies were
initially motivated by the theoretical connection between such models and
their four-dimensional analogues at high temperature, planar physics enjoys
nowadays the status of an interesting and self-contained topic in itself.
This position was achieved, in part, thanks to some peculiar features that
take place in this space-time, such as the coexistence of massive vector
gauge bosons and gauge invariance, and the possibility of having objects
displaying charge and statistical fractionization
\cite{frac,wilczek}. On the other hand, the interest in planar physical
models was also
remarkably motivated by Condensed Matter phenomena that
display planar dynamics. Among these ones, we may quote 
the Quantum Hall Effect \cite{qhe} and the High-Tc Superconductivity 
\cite {wilczek,laughlin}.\newline

Of particular interest is also the study of topological objects in this
framework. For example, topologically magnetic vortex-like solutions
naturally appear attached to electric charges whenever we are dealing with a
Chern-Simons-like electrodynamics (the so-called
Maxwell-Chern-Simons (MCS) model). In addition, it is well-known that this
composite entity (electric charge + magnetic vortex) may present anyonic
statistics thanks to the magnetic flux induced by the vortex 
\cite{wilczek,Jackiwvortex}.\\

Another sort of topological entities shows up whenever breaking 
Bianchi identity. These are generally characterized by a potential, 
$A_{\mu }$, which carries a singular structure. As it is well-known, 
such a kind of potentials first appeared in Dirac's paper on magnetic 
monopole \cite{dirac}. Actually, while in (3+1) dimensions the 
simplest solution appears like as a point-like magnetic monopole, we 
shall see that in the (2+1)-dimensional case, the breaking of the 
Bianchi identity leads us to a wider class of solutions, not 
restricted to magnetic ones (this is the reason why we call them 
Dirac-like objects).\newline

\begin{sloppypar}
Indeed, some works have dealt with such issues in both Euclidean \cite{HT,Pisarski} 
and Minkowskian \cite{prdtese} three-dimensional spaces. Here, it is 
worthy mentioning that the mass parameter was shown to be quantized in 
the Abelian version of the Maxwell-Chern-Simons model whenever 
Dirac-like monopoles interact with usual charges \cite{HT} (similarly 
to the result already known for theories whose gauge groups 
presented non-trivial third homotopy group \cite{djt}). In 
addition, classical and quantum consequences 
of the monopole potential acting upon a charged particle were recently analyzed
\cite{prdtese}.\newline\end{sloppypar}

In this article we wish to go further into this subject and investigate some
issues concerning the nature of such objects in three dimensions, as well as
some of their influences on the dynamics of particles. Then, in 
Section II we introduce a dimensional reduction of (3+1)D 
electrodynamics with magnetic sources to (2+1) dimensions. Such a 
presentation is interesting for highlighting the scalar nature of 
these sources in the planar case . Indeed, such a scheme yields  
two Abelian ``electrodynamics'' which do not 
have any explicit interplay between them. In addition, we point out 
the differences between these models, particularly in their magnetic 
sectors.\newline

Section III is devoted to the subject of the Dirac-like monopoles 
itself. There, we present a brief review of such objects  
introduced in Minkowskian and Euclidean spaces. Attention is given to 
the differences between them. We also present an analysis of 
the solutions admitted by the differential equation that shows up 
whenever Bianchi identity is broken in $(2+1)$ dimensions. \newline

In Section IV, we deal with the interaction between a Dirac-like 
monopole and a usual
particle. More precisely, our attention is focused on a 
Lorentz-violating non-minimal term, which couples monopole 
field-strength to neutral matter. Although violating 
Lorentz, it is shown to be invariant under CPT-symmetry. In addition, the 
equations of motion are similar to those we have for the case 
of a charged particle minimally interacting with the vector potential 
produced by a magnetic vortex. Indeed, by 
virtue of this similarity, such an interaction leads us to a 
Aharonov-Casher-like effect on the usual particle, produced by the 
tangential electric field of the monopole. \newline

Finally, our paper is closed by pointing out our Conclusions 
and Prospects for future investigation.\newline 
\section{The origin of the scalar nature of planar Dirac-like 
objects}

Here, we intend to give an alternative view of the scalar nature of the
Dirac-like monopoles in (2+1) dimensions. The proper study of the breaking
of the Bianchi identity in planar Abelian Maxwell and Maxwell-Chern-Simons
frameworks will be the goal of the next section, where we shall pay
attention, among others, to the tangential (azimuthal) behavior of the
electric-like field generated by a point-like ``magnetic source'' \cite
{Pisarski,prdtese}.\newline

In order to trace back the scalar nature of (2+1)D magnetic current to its
four-dimensional ancestor, we propose to carry out a plain dimensional
reduction of the (3+1)D Maxwell theory with electric ($j^{\hat{\mu}}$) and
magnetic ($k^{\hat{\mu}}$) sources, equations (\ref{dF4}-\ref{dFtil4}) below, to 
the planar case. Hereafter, we shall work in Minkowski space-time, but no 
difficult arises in carrying out a similar plain in the Euclidian 
case. \\

We start off from:\footnote{Our conventions read: 
$\hat{\mu},\hat{\nu},{\mbox{\rm etc}}=0,1,2,3$, ${\mbox{\rm 
diag}}\eta_{\hat{\mu}\hat{\nu}}=(+,-,-,-)$, and 
$\epsilon^{0123}=-\epsilon_{0123}=+1$. In addition, 
$\mu\nu,{\mbox{\rm etc}}=0,1,2$, ${\mbox{\rm 
diag}}\eta_{\mu\nu}=(+,-,-)$, and $\epsilon^{012}=\epsilon_{012}=+1$; 
while the planar spatial indices are labeled like as: $i,j=1,2$ and 
$\epsilon^{12}=\epsilon_{12}=+1$.} 
\begin{eqnarray}
& & \partial 
_{\hat{\mu}}F^{{\hat{\mu}}{\hat{\nu}}}=j^{\hat{\nu}} \,, \label{dF4} 
\\ &&\partial 
_{\hat{\mu}}\tilde{F}^{{\hat{\mu}}{\hat{\nu}}}=k^{\hat{\nu}}\,, 
\label{dFtil4} \end{eqnarray} with 
$F^{{\hat{\mu}}{\hat{\nu}}}=\partial 
^{\hat{\mu}}A^{\hat{\nu}}-\partial^{\hat{\nu}}A^{\hat{\mu}}$ and 
$\tilde{F}^{{\hat{\mu}}{\hat{\nu}}}=\frac{1}{2}\epsilon 
^{\hat{\mu}\hat{\nu}\hat{\kappa}\hat{\lambda}}F_{\hat{\kappa} 
\hat{\lambda}}$. \newline

First of all, we reduct the potential and currents like the `splitting' 
below:
\begin{eqnarray}
&&A^{\hat{\mu}}\quad {\longrightarrow}\quad
(A^{\mu};\;A^{3}\equiv S)\,,  \label{Asplitting} \\
&&j^{\hat{\mu}}\quad \longrightarrow \quad 
(j^{\mu};\;j^{3}\equiv\lambda)\,,  
\label{jsplitting} \\ &&k^{\hat{\mu}}\quad \longrightarrow \quad 
(k^{\mu};\;k^{3}\equiv\chi)\,.  \label{ksplitting} \end{eqnarray}  
Then, we realize that the (3+1) dimensional quantities are reducted to 
(2+1)D ones. For instance, $A^{\hat{\mu}}$ yields to a (2+1)D-vector, 
$A^\mu=(A^0,A^1,A^2)$, and to an {\em extra} scalar potential, $A^3\equiv 
S$. Notice, in addition, that from the point of view of a (2+1) dimensional 
frame the fields $A^\mu$ and $S$ are, at principle, completely independents 
(the same is valid for the currents). Similarly, $j^\mu$ and $k^\mu$ are 
the (2+1)D electric and magnetic currents, while $j^3\equiv\lambda$ and 
$k^3\equiv\chi$ represent the survivors of the 3rd components of the 
electric and magnetic genuine 4-currents, respectively.\newline

In addition, adopting the reduction \emph{ansatz} that the quantities 
do not depend on the 3rd-spatial coordinate, say, $\partial 
_{3}(f)=0$, where $f$ represents any potential or current, the 
field-strengths take the following forms after the dimensional reduction:
\begin{eqnarray} 
&&F^{{\hat{\mu}}{\hat{\nu}}}\quad \longrightarrow \quad 
(F^{\mu\nu};\;F^{\mu 3}\equiv G^{\mu})\,,  \label{Fsplitting} \\ 
&&\tilde{F}^{{\hat{\mu}}{\hat{\nu}}}\quad \longrightarrow \quad 
(\tilde{F}^{\mu \nu}\equiv\tilde{G}^{\mu\nu};\; 
\tilde{F}^{\mu 3}=\tilde{F}^{\mu}\,,  \label{Ftilsplitting} \end{eqnarray} 
where the new field-strengths are defined as: $F^{\mu \nu }=\partial ^{\mu 
}A^{\nu }-\partial ^{\nu }A^{\mu }$, $\tilde{F}^{\mu 
}=\frac{1}{2}\epsilon ^{\mu\nu\kappa}F_{\nu\kappa}$, $G^\mu=\partial^\mu S$ 
and $\tilde{G}^{\mu\nu}=\epsilon^{\mu\nu\kappa}G_\kappa$.\newline

Notice also that the usual planar electric and magnetic fields are
contained in the former field, say, $\tilde{F}^\mu= (-B; 
-\epsilon^{ij}E^j)$. In turn, the fields $G^\mu=\partial^\mu S$ and 
$\tilde{G}^{\mu\nu}= \epsilon^{\mu\nu\kappa}G_\kappa$ answer for the 
appearance of another ``electrodynamic'' model, like below.\newline

Now, taking into account relations (\ref
{jsplitting})-(\ref{Ftilsplitting}),  expressions (\ref{dF4}-\ref{dFtil4})
lead us to the two following
sets of equations: 
\begin{eqnarray}
& & \partial_\mu F^{\mu\nu}=j^\nu\quad {\mbox{\rm and}}\quad \partial_\mu
\tilde{F}^\mu=\chi \,,  \label{M2+1} \nn\\ \nn\\
& &\partial_\mu{G}^\mu=\lambda \quad{\mbox{\rm and}}\quad\partial_\mu
\tilde{G}^{\mu\nu}=k^\nu \,,\nn  \label{outro2+1}
\end{eqnarray}
from what we may still write down: 
\begin{eqnarray}
& & \left.\begin{array}{l}\epsilon^{ij}\partial^i B=\partial_t{E}^j+{j}^j\,,\\ 
\nabla\cdot\vec{E}=j^0=\rho\,,\\ \partial_t 
B+\epsilon^{ij}\partial^i{E}^j=\chi \,,\end{array}\right\} \label{planarelectromag} 
\\\nn\\
& & \left.\begin{array}{l}\epsilon^{ij}\partial^i b=\epsilon^{ij}
\partial_t{e}^i+{k}^j\,,\\ \epsilon^{ij}\partial^i{e}^j =k^0=\rho_m\,,\\
\partial_t b-\nabla\cdot\vec{e}=\lambda \,, \end{array}\right\} 
\label{planarmagelec} \end{eqnarray}
where the fields above are defined like below:
\ba
& & {E}^i\,=\,-\,\partial^i A^0-\partial_t {A}^i \qquad \mbox{and} 
\qquad B\,=\,\epsilon^{ij}\partial^i{A}^j\,,\nn\\\nn\\
& & {e}^i\,=\,-\,\partial^i S \qquad \mbox{and} \qquad 
b\,=\,\partial_t S\,.\nn 
\ea
Therefore, we realize that after dimensional reduction is 
implemented we get two independent electrodynamic-like models in (2+1) 
dimensions, each of them with its proper \emph{electric} and 
\emph{magnetic} sources. Indeed, the appearance of two non-coupled 
Abelian factor is nothing but a natural consequence of the 
reduction scheme. For instance, the latter one is equivalent to select 
the zero-mode sector of a more general dimensional reduction proposal, 
namely, the Kaluza-Klein \emph{ansatz} that relies on the compactness 
of the 3rd-spatial coordinate \cite{SS}. Thus, the natural 
$SO(2)$-symmetry associated to such a component is kept in 
(2+1)D, since the scalar field, $S$, is clearly invariant 
under rotations in the plane. We should also notice that the number of 
on-shell degrees of freedom is conserved in the reduction scheme. The 
two physical components of $A^{\hat{\mu}}$ lies, after dimensional 
reduction, in $A^{\mu }$ and in $S$, each of them carrying a unique degree 
of freedom.\newline

Furthermore, it is important to stress here that the breaking-down of the
Bianchi identity in (2+1)D and what we interpret as its associated magnetic
source in the planar world is the (2+1)D-manifestation of the 3rd component 
of the genuine magnetic 4-current\footnote{For this, notice that we are 
considering, as usually is done, the set of equations 
(\ref{planarelectromag}) as being the (2+1)-dimensional counterpart of 
the standard electrodynamics in 4 dimensions. The other Abelian 
sector, (\ref{planarmagelec}), that comes from the scalar potential, 
$S$, is then merely considered as being the partner of planar 
electromagnetism after the reduction procedure, even though the set 
(\ref {planarmagelec}) is the one that keeps the `genuine' 
(2+1)D reminiscent of the magnetic 4-current.}. This is how we 
understand the argument by Henneaux and Teitelboim \cite{HT}, that 
this charge rather behaves like an \emph{instanton} in the planar 
case. In addition, it is worthy noticing that $\chi$-charge is a 
pseudo-scalar, say, it changes its signal under parity: 
$\chi\to\chi^P=-\chi$, what is consistent with the equations of motion and 
with the fact that it appears, after dimensional reduction, as a 
reminiscent of the magnetic 4-current, which is a pseudo-vector. 
Similar behavior also occurs to all other currents and fields above. 
It would also be interesting to understand now, if possible to 
accomplish such a program, how the (3+1)-dimensional Dirac 
quantization condition may induce an analogue on the $\chi 
$-charge.\newline

So, our claim is that, once we start with a 4D Maxwell theory enriched
by the presence of magnetic monopoles and, if some physical system is
considered such that non-planar effects are negligible in comparison with planar 
effects, such a system may reveal particles that interact via two quantum 
numbers and one of them may induce an electric field with azimuthal 
configuration (see Section III, for details).\newline

\section{Analyzing the breaking of Bianchi identity}

Dirac-like objects come about through breaking Bianchi identity, as we have
already mentioned. In $(3+1)$ dimensions, when we consider Maxwell
electrodynamics with magnetic sources, we have the following equations $
\partial_{\hat{\mu}} F^{\hat{\mu}\hat{\nu}}=j^{\hat{\nu}}$ and
$\partial_{\hat{\mu}} \tilde{F}^{\hat{\mu}\hat{\nu}}=k^{\hat{\nu}}$. 
There, the magnetic Gauss law, $\nabla\cdot\vec{B}=\chi^0$, whenever 
taken for a point-like source, $\chi^0=g\delta^3(\vec{x})$, leads us 
to the concept of a genuine magnetic monopole since 
$\vec{B}=g\vec{x}/4\pi\,|\vec{x}|^3$, in analogy to the electric field 
produced by an isolated point-like electric charge. Clearly, such a 
similarity takes place because of the duality between electric and 
magnetic sectors, say, $\vec{E}$ and $\vec{B}$ are rank-1 tensors 
(notice that this happens only in 4 dimensions!).\\

On the other hand, when considered in $(2+1)$ dimensions, the broken
version of Bianchi identity yields to 
\ba\label{9}
\partial^\mu\tilde{F}_\mu=\partial_t{B}+\epsilon^{ij}\partial^i{E}^j= 
\chi \,. 
\ea
Here, there is no Gauss law for the magnetic field, which implies, in turn, 
that magnetic monopoles like as those we encountered in 4 dimensions, are 
no longer present. Thus, although rising up like as genuine magnetic 
sources in $(3+1)$-dimensional electromagnetism, the present objects are 
expected to exhibit several differences whenever compared to the first 
ones.\newline

Furthermore, in dealing with the massless case the breaking of the Bianchi
identity causes no
effect on the equations of motion, i.e., electric current is automatically
conserved,
\be
\label{dFMCS0}
\partial_\nu\partial_\mu F^{\mu\nu}=\partial_\nu j^\nu=0\;\;.
\ee
Nevertheless, when the Chern-Simons term,
${\cal L}_{CS}=mA_\mu\tilde{F}^\mu$, is taken into account, things change 
deeply. Now, the equations of motion acquire an extra (topological) current 
term, \begin{equation}\label{dFMCS}
\partial_\mu F^{\mu\nu}=j^\nu+m\tilde{F}^\nu\;\;,
\end{equation}
which yields to $\nabla\cdot\vec{E}=\rho+mB$ and \
$\epsilon^{ij}\partial^i B=\partial_t 
E^j+\,{j}^j\,+\,m\epsilon^{ij}{E}^i$.\\
 
Now, contrary to the massless case, if Dirac-like objects are
introduced, $\partial^\mu\tilde{F}_\mu=\chi$, then current is no longer
conserved, say: 
\begin{equation}
\partial_\nu\partial_\mu F^{\mu\nu}=\,m\,\chi\,,
\end{equation}
and gauge symmetry is lost. Thus, in order to restore such a symmetry we
should suppose that the appearance of Dirac-like entities naturally induces an
extra electric current, 
\begin{equation}  
\label{101}
j^\nu_M=-m\tilde{F}^\nu\;\;,
\end{equation}
so that equation (\ref{dFMCS}) is modified to 
\begin{equation}  \label{12}
\partial_\mu F^{\mu\nu}=J^\nu +m\tilde{F}^\nu
\end{equation}
and it is now, identically conserved 
\begin{equation}  \label{14}
\partial_\nu\partial_\mu F^{\mu\nu}=\partial_\nu J^\nu+m\partial_{\nu}\tilde{F}^\nu=0\,,
\end{equation}
where $J^\nu=j^\nu+j^\nu_M$ is the total (usual + topologically induced)
electric current (for further details, see Refs. \cite{HT,Pisarski}).\newline

On the other hand, in the 3-dimensional Euclidean space, we have that: 
\begin{equation}
\partial_\mu\tilde{F}^\mu=\partial_\tau\tilde{F}^0 +\partial_i 
\tilde{F}^i=\chi\,, \;\;. \end{equation}
Now, writing $\tilde{F}^\mu=-\partial^\mu \phi$,
and taking $\chi$ as being a point, $\chi=g\,\delta^3(\vec{x})$, we get
$\partial^2\phi= -g\,\delta^3(\vec{x})$,
whose solution reads 
\begin{equation}
\phi(\vec{x})\,=-g/4\pi\,|\vec{x}|\;\;,
\end{equation}
where $|\vec{x}|=\sqrt{\tau^2+x_i^2}\;\;$.\\

The fields, in turn, are given by
$\tilde{F}^\mu=-\partial^\mu \phi= g x^\mu/4\pi\, |\vec{x}|^3$,
or still (let us recall that $\tilde{F}^\mu=(-B;-\epsilon^{ij}E^j)$):
\begin{eqnarray}
B=-\frac{g}{4\pi}\frac{x^0}{|\vec{x}|^3}\quad{\mbox{\rm and}}\quad 
E_i= -\frac{g}{4\pi}\frac{\epsilon_{ij}x_j}{|\vec{x}|^3}\,,  
\label{EBEucli} \end{eqnarray} what clearly shows us that genuine 
magnetic monopoles, whose only effect is the production of a magnetic 
field, as we realized in 4 dimensions, no longer take place here.\\

It is particularly noticeable the tangential character of the electric 
field above (like as its analogue in Minkowski case, below), in 
contrast to what we expect from usual electric or even magnetic poles. 
When working in the MCS framework, the induced electric current, 
equation (\ref{101}), is readily found to be 
\begin{equation}
\rho_M=-\frac{mg}{4\pi}\frac{\tau}{|\vec{x}|^3}\quad{\mbox{\rm
and}}\quad j^i_M=-\frac{mg}{4\pi}\frac{x^i}{|\vec{x}|^3}\;\;,
\end{equation}
which presents radial-like dependence. Further details about such a subject,
including investigation on non-Abelian versions of such entities may be
found in Ref. \cite{Pisarski}.\newline

Now, let us return to Minkowski space-time and let us analyze the 
structure and solutions of equation (\ref{9}) in details. Rewriting this 
equation in components we get (hereafter, we shall use a subscript $g$
in order to distinguish these fields from the usual electric and magnetic
ones and from their Euclidean counterparts), 
\begin{eqnarray}  \label{BidentEB}
\partial_t B_g\,+\epsilon^{ij}\partial^i{E}^{j}_g=\chi\;\;.
\end{eqnarray}
whose point-like solutions may be obtained by considering special 
situations. First, considering the \emph{static limit of the 
fields} in equation above, we obtain 
\begin{equation}  \label{27}
\epsilon^{ij}\partial^i{E}^j_g\,=\,g\,\delta^2\,(x)\;\;,
\end{equation}
which, when written in terms of the potential 
$\vec{E}_g=-\nabla\Phi_g$ gets the following form: 
\begin{equation}
\left[ \partial_x,\partial_y \right]\Phi_g\,=\,-\,g\delta^2(\vec{x})\;\;,
\end{equation}
whose solution reads (with $r=|\vec{r}|=\sqrt{x^2+y^2}$ and 
$\theta=\arctan(y/x)$, as usual)
\begin{eqnarray}
\Phi_g(\vec{x})\,=\,-\,\frac{g}{2\pi}\arctan\left(\frac{y}{x}\right)\,
= \,-\,\frac{g}{2\pi}\theta\;\;. \end{eqnarray}
Notice the remarkable feature of such a potential: it has angular rather
than radial dependence. Notice also its singular structure: the
angle-function is not well-defined at the origin, like the string-like
presented by the vector potential associated to a genuine magnetic monopole 
in (3+1) dimensions. Besides, it is a multivalued function and the 
corresponding electric field (see below) is not a conservative one, fact 
already indicated by equation (\ref{27}). Indeed, its associated electric 
field reads: \begin{eqnarray}  \label{Emon}
\vec{E}_g\,=\frac{g}{2\pi}\frac{x\hat{j}-y\hat{i}}{x^2+y^2}\,=
\frac{g}{2\pi}\frac{\hat{e}_\theta}{r}\;\;,
\end{eqnarray}
like as in Ref.\cite{prdtese}, which has an {\em azimuthal} rather than 
a radial-like vector behavior. In addition, by demanding null 
radiation at this static limit, $\int_V \partial_i\epsilon_{ij}E_j B\, 
dV=0$, it readily follows that in this case $B_g$ must vanish. 
Therefore this static solution appears much 
as due to a peculiar topological electric charge, rather than to a 
magnetic monopole.\\

Furthermore, if we compare it with the vector potential associated to 
a magnetic vortex ($\Phi_B$ being its magnetic flux), 
\begin{equation}
\vec{A}_v(\vec{x})\,=\,\frac{\Phi_B}{2\,\pi\,r}\,\hat{e}_\theta
\end{equation}
we may identify a kind of ``duality'' between them. Actually, the 
magnetic vortex may be obtained from a Dirac-like monopole, equation 
(\ref{Emon}), by interchanging the vectors $\vA_v$ and $\vE_g$, 
together with $g$ and $\Phi_B$ (let us recall that in the case of a  
usual electric charge a similar identification requires the 
interchanging between $\vA_v$ and the dual of $\vE$).\\

Before carrying on the analysis of other possible solutions, let us pay
attention to the (topological) electric current induced by the appearance of
this monopole in the MCS framework. From equations (\ref{101}) and 
(\ref{Emon}) it follows that \begin{eqnarray}
\rho_M=0 \qquad \mbox{\rm and} \qquad j^i_M=\frac{mg}{2\pi} 
\frac{x^i}{|\vec{x}|^2}\;\;. \end{eqnarray}
which is radial, and implies that equation (\ref{14}) is satisfied.\newline

The second situation is the \emph{radial-like electric field}. Now,
searching for solutions of equation (\ref{BidentEB}) that present 
$\epsilon^{ij}\partial^i{E}^j=0$, we are left with
\ba
\partial_t B_g=\chi\,.
\ea 
Here, let us take the simplest time-dependent configuration for 
$\chi$-charge, $\chi=g\delta(t)\delta^2(\vec{x})$, which is similar to 
the one we have taken in Euclidean space. Such a case is readily solved 
by taking: \begin{eqnarray}  \label{Bvortex} 
B_g(\vec{x},t)=g\delta^2(\vec{x})\Theta(t)\,, \end{eqnarray}
what is clearly the magnetic field due to a vortex-like object with flux
equal to $2\pi g$ (created at $t=0$).\\

We may also think about a configuration which reverses the 
direction of the magnetic flux, say 
\begin{equation}
B_g(\vec{x},t)=(g/2)\,\delta^2(\vec{x})[\Theta(t)-\Theta(-t)]\;\;,
\end{equation}
which clearly represents a magnetic vortex with flux $-g/2$ that 
changes its signal at $t=0$, or still, the destruction of a 
$-g/2$-flux vortex at $t=0$ with the simultaneous creation of another 
one with flux $g/2$.\\ 

For such an object the equation (\ref{Bvortex}), its topologically induced 
electric current, takes the form \begin{equation}
\rho_M=mg\delta^2(\vec{x})\Theta(t)\quad{\mbox{\rm and}}\quad j^i_M=0\;\;,
\end{equation}
which represents a point-like electric charge of strength $-mg$ created at
$t=0$. In addition, since $B$ and $\rho_M$ above are located at 
$\vec{x}=0$, we conclude that in MCS case, the appearance of a 
composite vortex-electric charge may be alternatively provided through 
the introduction of a vortex-like solution, like as (\ref{Bvortex}), 
whenever breaking the Bianchi identity.\newline

A {\em more general solution} associated to equation (\ref{BidentEB}) is 
obtained by taking a ``mixture'' of previous ones. Let be 
$\chi=g\delta^2(\vec{x})\delta(t)$ and let us combine previous solutions, 
like below: \begin{eqnarray}  \label{solgeneral}
B_g(\vec{x},t)= \frac{g}{2}\delta^2(\vec{x})\Theta(t)\quad{\mbox{\rm
and}} \qquad \vec{E}_g=+\frac{g}{4\pi}\frac{\hat{e}_\theta}{|\vec{x}|}%
\delta(t)\;\;.
\end{eqnarray}
As it is clear, such expressions take together the solutions associated to
the vortex-like, created at $t=0$, and to the Dirac-like monopole, 
only at $t=0$. The electric field above induces \begin{equation}
\vec{j}_M=\frac{mg}{4\pi}\frac{\vec{x}}{|\vec{x}|^2}\delta(t)\;\;,
\end{equation}
which takes electric charges away from the origin at $t=0$, while 
\begin{equation}
\rho_M=\frac{mg}{2}\delta^2(\vec{x})\Theta(t)\;\;,
\end{equation}
corresponds to the induced charge at $\vec{x}=0$, provided by
$\vec{j}_M$.\newline

Let us compare solution above with that we have in Euclidean space, equation (\ref
{EBEucli}). Monopole-like solution in Euclidean space, equation (\ref{EBEucli})
represents an object that produces a tangential electric field and a
`radial-like' magnetic field, both of them proportional to 
$1/|\vec{x}|^{2}$. On the other hand, if we consider one of its 
analogue in Minkowskian space-time, solution (\ref{solgeneral}), we 
realize that in this case the monopole-like solution gives us a 
magnetic field confined to a point in space, a vortex, and a 
tangentially directed electric field which is proportional to 
$1/|\vec{x}|$ and, in addition, takes place only at $t=0$. Therefore, 
we conclude that the dimension and structure (topology, etc.) of the 
space-time is decisive for the solutions of the fields associated to 
Dirac-like objects.\newline

Before close this section, let us pay attention to the issue 
concerning the introduction of such entities in electrodynamic-like 
models, namely, three-dimensional Abelian gauge theories. First of 
all, notice that in the Bianchi identity breaking scenario, no space is 
reserved to the appearance of a mass term, say, we could not provide a 
mass gap for the radiation associated to the monopole-like field (for the 
time being, we are supposing different radiation for dynamical and 
geometrical sectors of the equations of motion).\newline

Now, in the case of the pure Maxwell (massless) model, the breaking of
Bianchi identity causes no additional trouble in the dynamical sector, for
instance, electric current remains conserved. Therefore, in this case,
nothing prevent us from taking into account that we have indeed a unique
(massless) radiation which mediates the interaction among usual electric
charges (usual electric and magnetic fields), among Dirac-like objects 
($\vec{E}_g$ and $B_g$), and also among the first and the latter ones. 
It is worthy noticing that such an identification of apparently 
distinct sorts of interaction as being manifestation of only one kind 
of radiation is possible here because of all the required potentials 
and fields are gapless.\newline

However, if we try to apply a similar identification in the MCS framework we
meet serious troubles. Here, usual radiation is naturally massive. For
example, the electric field between two static electric charges is
proportional to $K_{0}(m|\vec{x}|)$ (with $K_{0}$ being the modified Bessel
function of $2nd$ kind at $0th$ order), and so it is a short-range
interaction. In deep contrast, the tangential electric field due to a
monopole-like solution carries no hint about mass, see equations (\ref{EBEucli})
and (\ref{Emon}). Actually, as far as we have tried, no way was found in
order to identify both types of interaction as produced by the same
radiation. This would require an action which has already 
enclosed usual and monopole-like potentials as its basic ingredients, 
and so, answer whether is required one or two kinds of radiation.\newline 
\begin{sloppypar}\section{Neutral particles non-minimally 
coupled to monopole field and the Aharonov-Casher 
effect}\end{sloppypar}

In this section we shall consider a non-minimal coupling of a spinor 
field with the electric field generated by a `static monopole', equation 
(\ref{Emon}). First, however, we shortly review some basic aspects 
of the usual non-minimal case, mainly those concerning the 
Aharonov-Casher effect. Indeed, it is a peculiarity of (2+1) 
dimensions that even spinless particles may carry anomalous magnetic 
momentum, whenever interacting with an electromagnetic field. This 
lies in the fact that the momentum may be naturally supplemented by 
the dual field-strength, say:
\ba
\partial_\mu\;\longrightarrow\;\partial_\mu +i\,h\,\tilde{F}_\mu\,,
\ea
where $h$ measures the planar anomalous magnetic momentum of the 
matter (see, for example, Refs.\cite{nonminimal,CK}, for further 
details).\\

Now, let us take the electromagnetic field, 
$\tilde{F}_\mu=(-B;\epsilon_{ij}E_j)$, produced by a usual point-like electric 
charge, say, $B=0$ and $\vec{E}=q\vec{x}/2\pi|\vec{x}|^2$. Then, if we 
consider the interaction of such a field with a given particle (mass $m$), we 
find that the energy-operator of the latter reads \ba
H=\frac{1}{2m}(\partial_i 
+i\,h\,\tilde{F}_i)^2=\frac{1}{2m}(\partial_i +i\,h\,\epsilon_{ij} 
E_j)^2 \,.\label{H1} \ea
In addition, if the ``free'' wave-functions associated to the particle 
satisfy
\ba
\left(i\partial_0+\frac{1}{2m}\nabla^2\right)\psi^{(0)}=0\,, 
\label{Schfree} \ea
then, the WKB approximation yields the new functions:
\ba
\psi(\vx,t)=\psi^{(0)}(\vx,t)\,{\mbox{\rm 
exp}}\left[-i\int\,dx_\mu\,h\,\tilde{F}^\mu\right]\,.\label{WKB1} 
\ea
Thus, we realize that, the addition of the field $\tilde{F}^\mu$ to the usual 
momentum is equivalent to introduce (at WKB-level) a non-integrable phase to 
the former wave-functions. Clearly, a similar plain also holds in the 
case of the minimal coupling, 
$\partial_\mu\;\to\;\partial_\mu+ieA_\mu$ (see Ref. \cite{CK}), which 
is responsible for the appearance of the so-celebrated Aharonov-Bohm 
(AB) effect \cite{AB} and, in (2+1) dimensions magnetic flux-carrying
particles leads to fractional statistics \cite{wilczek}.\\

The interesting point to be noticed here is that, if we consider 
the particle performs a spatial loop, say $\theta$, around the charge 
$q$, then
\ba
\theta=h\,\oint\,dl_i\tilde{F}^i=h\,\oint\,dl_i 
\epsilon^{ij}E^j=\frac{hq}{2\pi}\oint\,dl_i\epsilon^{ij} 
\frac{x^j}{|\vx|^2}\,.\label{alfa1}
\ea  
Now, since $dl_i=\epsilon_{ij}dx_j$ ($d\vx$ is radial) and 
$\nabla\cdot\vx/|\vx|^2\,=\,2\pi\delta^2(\vx)$ we finally obtain:
\ba
\theta=\frac{hq}{2\pi}\int_S\, dS 
\nabla\cdot\frac{\vx}{|\vx|^2}=hq\,.\label{alpha2} 
\ea
Therefore, we have that $\psi(\vx,t)=\psi^{(0)}(\vx,t)\,e^{i\theta}$, 
where $\theta$ is the Aharonov-Casher (AC) phase provided by the 
electric field $\vec{E}=q\vx/2\pi|\vx|^2$ (for further details see 
Refs. \cite{CK,AC,hagen2} and related references therein).\\

In our present case, the counterpart of the electric field above reads 
like equation (\ref{Emon}), $\vec{E}_g=g\,(x\hat{j}-y\hat{i})/2\pi|\vx|^2$. 
Then, $\tilde{F}^i_g=-\epsilon^{ij}E^j_g=g\vx/2\pi|\vx|^2$ is already 
radial. In this case, a similar loop as in the previous case, 
$\theta'$, vanishes:
\ba
\theta'=h\,\oint\,dl_i\tilde{F}^i=\frac{hg}{2\pi}\oint\,dl_i 
\frac{x^i}{|\vx|^2}=0\,. 
\ea
Then, our monopole does not induce an AC-phase on a given particle 
if they interact in the usual non-minimal way. In addition, we should 
notice that the contrast between the cases above comes from the fact 
that, in the first one, say, $\vec{E}=q\vx/2\pi|\vx|^2$, the dual operation 
induced whenever taking $\tilde{F}^i=-\epsilon^{ij}E^j$ is exactly 
compensated by an extra one associated to $dl_i=\epsilon_{ij}dx_j$.\\

In view of such an aspect, we shall consider here the (Lorentz-odd) 
non-minimal term like below (coupled, for concreteness, to spinors):
\ba
{\cal L}'=\overline{\psi}(i\partial_\mu\gamma^\mu 
-M +ia\gamma^0\gamma^\mu\tilde{F}_\mu)\psi\label{Llinha}\,,
\ea
whose equation of motion reads:
\ba
(i\partial_\mu\gamma^\mu 
-M +ia\gamma^0\gamma^\mu\tilde{F}_\mu)\psi=0
\ea
Before studying some properties of such a term, like as its connection 
with AC effect, we shall give attention to its behavior under 
special properties, say, gauge invariance, Charge Conjugation (C), Parity (P) 
and Time Reversal (T). For this, let us take $\gamma^0=\sigma^z$, 
$\gamma^1=i\sigma^x$ and $\gamma^2=i\sigma^y$ as the representation of the Dirac 
matrices in $(2+1)$-dimensions.\\

First, analyzing the behavior of ${\cal L}'$  under gauge transformations, we 
may clearly realize its gauge-invariance, since 
\bn\left.\begin{array}{l}
\delta \bar{\psi} =\epsilon\,\bar{\psi}\;\;, \\
\delta \psi =-\,\epsilon\,\psi\;\;, \\
\delta E_{x} =\epsilon \gamma^1 \eta\;\;\\
\delta E_{y} =2 \epsilon \gamma^2 \eta\;\;, \\
\delta B =\epsilon \eta\;\;,\end{array}\right\}
\en
where $\epsilon$ is a global gauge parameter and $\eta$ is a local auxiliary 
field which helps in the gauge invariance.\\

On the other hand, using the identity 
$\gamma^\mu\gamma^\nu=\eta^{\mu\nu}-i\epsilon^{\mu\nu\kappa}\gamma_\kappa$, 
we may write:
\ba
ia\overline{\psi}\gamma^0\gamma^\mu\tilde{F}_\mu\psi= 
-ia\,B\,\overline{\psi}\psi 
+a\overline{\psi}\vec{\gamma}\cdot\vec{E}\psi\label{componentes} \,,
\ea                                                                                                                                                         
Now, let us see how the terms above behave under $C$, $P$ and $T$ operations. 
Let us strat off from:
\ba
& & ia\,B\,\overline{\psi}\psi\; 
\stackrel{C}{\longrightarrow}\;-ia\,B\, \overline{\psi}\psi 
\,,\label{C1}\\ 
& & ia\,B\,\overline{\psi}\psi\; 
\stackrel{P}{\longrightarrow}\;-ia\,B\,\overline{\psi}\psi 
\,,\label{P1}\\ 
& & 
ia\,B\,\overline{\psi}\psi\;\stackrel{T}{\longrightarrow}\;+ia\,B\, 
\overline{\psi}\psi \,,\label{T1} \ea
then, although breaking $C$ and $P$, such  a term keeps $T$-invariance 
and so $CPT$-symmetry is preserved. In addition, let us notice that 
this term provides an extra (imaginary) mass for the fermions when $B\neq 0$ 
(while its usual counterpart, $fB\overline{\psi}\gamma^0\psi$, couples 
to the electric charge)\footnote{Then, in view of its 
imaginary nature, we should take it away from equation (\ref{Llinha}) 
in order to maintain the real characterof this Lagrangian. This is done in 
what follows (see eq.(\ref{Ltrue}), and related discussion).}.\\

On the other hand, the spatial components behave like follows:
\ba
& & a\overline{\psi}\vec{\gamma}\cdot\vec{E}\psi\;\stackrel{C} 
{\longrightarrow}\;+a\overline{\psi}\vec{\gamma}\cdot\vec{E}\psi 
\,,\label{C2}\\
& & a\overline{\psi}\vec{\gamma}\cdot\vec{E}\psi\;\stackrel{P} 
{\longrightarrow}\;+a\overline{\psi}\vec{\gamma}\cdot\vec{E}\psi 
\,,\label{P2}\\
& & a\overline{\psi}\vec{\gamma}\cdot\vec{E}\psi\;\stackrel{T} 
{\longrightarrow}\;+a\overline{\psi}\vec{\gamma}\cdot\vec{E}\psi 
\,,\label{T2}
\ea 
which state us that the term above, coupling the current to the 
electric field, preserves all of these symmetries above, and CPT is 
obviously kept. Here, it should be noted that its usual counterpart, 
$f\overline{\psi}\gamma^i\epsilon^{ij}E^j\psi$, which couples the 
current density to the dual electric field is $P$ and $T$-odd (while 
respects $CPT$, since it is $C$-even). Then, when Lorentz and 
CPT-symmetries are taken into account, we recognize profound 
differences between the present and the usual non-minimal couplings. 
Thus, our proposal may be viewed as a low-energy alternative to the 
standard term, particularly in those cases in which neither $P$ nor
$T$ operation is broken.\\

Hereafter we shall focus our attention to the 
field produced by the monopole, equation (\ref{Emon}), and its consequences 
concerning AC phase as well. Thus, we shall work with the 
(Lorentz-violating) Lagrangian below:
\ba
{\cal L}=\overline{\psi}(i\partial_t\gamma^0 -i\partial_i\gamma^i
-M +ia\gamma^0\gamma^i\tilde{F}_i)\psi\label{Ltrue}\,,
\ea
which leads to the following eqs. of motion:
\ba
i \partial_{t} \psi(x)= [\gamma_{0} \vec{\gamma}\cdot(-i \vec{\nabla} + a 
\vec{E}_g) + M \gamma_{0}\,]\,\psi(x).\label{eqpsi} \ea 
In addition, it is easy to show that Lagrangian(\ref{Ltrue}) is invariant under 
gauge transformations, say:
\bn \left.\begin{array}{l}
\delta \bar{\psi}= \epsilon \bar{\psi}\;\;,\\
\delta \psi = -\epsilon \psi\;\;, \\
\delta E_x = \epsilon \gamma^2 \phi \;\;, \\
\delta E_y = \epsilon \gamma^1 \phi\end{array}\right\}
\en
where $\phi$ is a local auxiliary field analog to the $\eta$ field described 
above.\\

Now, taking the field generated by a point-like monopole, 
$E_i=(E_g)_i=g\epsilon_{ij}x_j/2\pi|\vx|^2$, and working in the WKB 
approximation, we have that:
\ba
\psi(\vx,t)=\psi^{(0)}(\vx,t) {\mbox{\rm 
exp}}\left[-i\,a\int\,dl_i\,E^i_g\right]\,,\label{WKBpsi} 
\ea
with $\psi^{(0)}$ satisfying equation(\ref{eqpsi}) when $\vec{E}_g$ is 
vanishing. Now, supposing that the fermion perform a loop, $\alpha_g$, 
around the monopole,
\ba
\alpha_g=a\oint\,dl_iE^i_g=\frac{ag}{2\pi}\oint\,\epsilon_{ij}dx_j
\epsilon^{ik}E^k_g=ag\,,\label{ACphase}
\ea
which clearly represents the AC phase on the fermion wave-function 
produced by the monopole field, $\vec{E}_g$. We should stress, once 
more, that such a phase comes from a $C$, $P$ and $T$-invariant 
non-minimal coupling.\\

In this way we have carried out the duality symmetry found at the previous 
section between the electric field produced by a static monopole and the vector 
potential of a magnetic vortex to the level of quantum mechanics.\\

\begin{sloppypar}Now, in order to study the behavior of the wave-functions it is 
more convenient to work with the second-order differential equation in polar 
coordinates ($x=r\cos\varphi$ and $y=r\sin\varphi$), like 
follows:\end{sloppypar}

\ba\label{rfi}
\left[ \frac{1}{r}\frac{\partial }{\partial r}\left( r\frac{\partial} 
{\partial r}\right) +\frac{1}{r^{2}}\left( \frac{\partial }{\partial 
\varphi} +i\alpha \right) ^{2}+\right.\nn\hspace{2cm}\\
\left.-\alpha s \sigma _{z}\frac{1}{r}\delta 
(r)+k^{2}\right] \psi (r,\phi )=0\,, \ea

\noindent where $\alpha =\frac{ag}{2\pi }$, 
$k^{2}=E^{2}- M^{2}$, and we are using $s= +1$ for ``spin up" and $s=-1$ for 
``spin up" (the actual spin of the spinors is $s/2$). \\

This equation is well known from the Aharonov-Bohm (AB) effect for
relativistic particles. In fact, this is a sort of Aharonov-Casher (AC)
effect since
we have a neutral particle in the presence of an electric 
field, as we have discussed above. Furthermore, the presence of spin leads to 
the $\delta (r)$ interaction, which mimics the interaction of the spin of the 
particle with a magnetic vortex (Zeeman effect), and may be interpreted as a 
contact interaction of the spin with the monopole itself. Although this residual 
interaction term vanishes outside the location of the monopole, the influence of 
the monopole on the dynamics of the particles is still felt by the induction of 
a non-trivial phase, equations (\ref{WKBpsi}-\ref{ACphase}) and, consequently, 
on the phase shift of the scattered wave-function.\\

In the works of Refs.\cite{hagen1, hagen2, china} such a problem is treated in 
the context of the AB effect. The authors adopt different approaches to 
regularize the delta function potential, which in our case is equivalent to 
suppose that the radius $R$ of the monopole is finite and is taken to 
zero at the end of the calculations.\\ 

To quote the main results, we consider the upper component of $\psi (r,\varphi 
)$ and expand it as

\begin{equation}
\psi _{1}=\sum\limits_{-\infty }^{m=+\infty }f_{m}\,(r)\,e^{im\varphi}\;\;,
\end{equation}

\noindent where $f_{m}(r)$ obeys the following equation

\ba
\left[\frac{1}{r}\frac{d}{dr}\left( r\frac{d} 
{dr}\right) -\frac{(m+\alpha )^{2}}{r^{2}}+\right.\nn\hspace{3cm}\\
\left.-\frac{\alpha 
}{R}\delta (r-R)+k^{2}\right] f_{m}(r)=0\,, \ea

\noindent with the following boundary conditions
\ba\label{f}\left.\begin{array}{l}
f_{m}(R-\varepsilon )=f_{m}(R+\varepsilon )\,,\\ \\
R\frac{d}{dr}f_{m}|_{R-\varepsilon }^{R+\varepsilon }=\alpha f_{m}(R)\,,
\end{array}\right\}
\ea

\noindent which incorporate the effect of the delta function.\\

By writing the $f_{m}(r)$ in terms of Bessel functions

\[
f_{m}(r)=\left\{ 
\begin{array}{c}
A_{m}J_{|m+\alpha |}(k \,r)+B_{m}J_{-|m+\alpha |}(k 
\,r)\quad r>R \\\nn\\ 
C_{m}J_{m}(k \,r)\qquad \qquad \qquad \qquad r<R
\end{array}
\right. 
\]

\noindent and using equations (\ref{f}) to determine the coefficients of the 
Bessel functions this renormalization method allows for the irregular function 
$J_{-|m+\alpha |}$ to contribute if the following relations

\[ |m| + |m + \alpha|= - \alpha s \]
and
\[ |m| + \alpha s + 1 > 0 ,\]

\noindent are simultaneously satisfied \cite{hagen1, hagen2} . \\

The same kind of problem was analyzed in \cite{gerbert} by using the 
self-adjoint method and found to be equivalent to the renormalization method if 
some relations between the self-adjoint parameter and the renormalized coupling 
constant are satisfied \cite{jackiw2}.\\

\section{Conclusions and Prospects}

In the present paper, attention was given to Dirac-like monopoles in
three-dimensional Abelian Maxwell and Maxwell-Chern-Simons models.
Initially, we gave an alternative view on the scalar nature of such objects
in planar world. This was done by carrying out the dimensional reduction of
four-dimensional Maxwell theory, enriched by magnetic sources, to three
dimensions. There, we realized the appearance of two independent Abelian
factors, one related to the usual $A^\mu$-potential ($\vec{E}$ and $B$
fields), while the another is implemented by a scalar potential, $S$. 
In addition, we have also verified that the broken Bianchi identity of 
the $A^\mu$-sector (usual planar electromagnetism), $\partial_\mu 
\tilde{F}^\mu=\chi$, presents a pseudo-scalar that is the survivor of the 3rd 
component of the genuine magnetic 4-current.\newline

Furthermore, in analyzing the structure of the solutions of $\partial 
_{\mu} \tilde{F}^{\mu }=\chi $, we have realized it admits a wider 
class of solutions than so far considered in the literature. Indeed, 
in Minkowski space-time, we have seen that not only the azimuthal-like 
electric field shows up, but also, magnetic vortex-like solutions may 
appear as well.\\

In addition, when neutral matter interacts non-minimally with Dirac-like
monopoles in a particular way, say, via $i\gamma ^{0}\gamma^i 
\tilde{F}_i$, then an analogous to the Aharonov-Casher effect is 
exhibited by such particles. We have also found 
some subtleties which have to be taken into account carefully because 
of the effective delta function potential whose origin rests on the 
``contact'' interaction between the particle and the monopole. This is 
still under study as well as the consequences of the allowed solutions 
on the angular momentum of the particle and perhaps on the 
quantization of the parameter $\alpha .$ \\

Before pointing out our Prospects, we would like to take once more the issue
concerning the Bianchi identity in (2+1) dimensions. First of all, let us
suppose it holds. Now, let us consider a physical system in which magnetic
vortex are created. For instance, when the external magnetic field is
suitably increased in high-Tc superconductors samples. More precisely, let
us imagine one vortex is created at $t=t_{1}$ and at spatial origin. Then,
the superconductor is supplemented by 
$B_{1}(\vec{x},t)=b_{1}\delta^{2} (\vec{x})\Theta (t-t_{1})$. On the 
other hand, since $\partial _{\mu }\tilde{F}^{\mu }=\partial 
_{t}B-\nabla \wedge \vec{E}=0$ then $B_{1}$ above must induce a 
tangential-like electric field, $\vec{E}_{1}=b_{1}\frac{\hat{e} 
_{\theta }}{r}\delta (t-t_{1})$, in order to prevent the breaking of 
Bianchi identity. Hence, we conclude that the azimuthal-like electric 
field may appear even in standard planar electromagnetism, say, 
without Dirac-like objects. Actually, $\vec{E}_{1}$ above survived 
only at $t=t_{1}$ because we have supposed that the creation of the 
vortex is also instantaneous. Now, if a finite time is needed for 
creating such a vortex, then we expect that this electric field will 
also take place during all this time.\newline

As perspectives for future investigation, we may quote, among others, the
issue concerning the effect of the dimensional reduction on the so-called
Dirac quantization condition in (3+1) dimensions and which would be its
counterpart in the planar world ( as far as we have understood, the
Henneaux-Teitelboim condition \cite{HT} does not answer for such a
point, since it seems to be valid only when the topological mass is 
non-vanishing).\newline

The relevance of the 
scalar field, $S$, appearing in Section 2 is also under investigation in the context of the 
so-called {\em statistical field}. Actually, by taking an Abelian 
Lagrangian which contains the usual Maxwell and the $\theta$-term as 
well, $\theta\tilde{F}_{\hat{\mu}\hat{\nu}}F^{\hat{\mu}\hat{\nu}}$, in 
(3+1) dimensions, we have seen that, after a suitable dimensional 
reduction scheme, we naturally generate, in (2+1) dimensions, a model 
which encloses the kinetic terms for $A_\mu$ and $S$ as well as 
another one that links both of these fields by means of a 
Chern-Simons-like term. Indeed, by identifying $a_\mu=\partial_\mu S$, 
we clearly realize that such a subsequent model is actually that for 
the (non-dynamical) statistical field, $a_\mu$, in which this field 
enters in order to restore Parity symmetry. Further results will 
appear elsewhere \cite{workinprogress}.\\   

The discussion raised up in the preceding paragraphs may also lead us 
to interesting results, in particular, for providing a link between 
fundamental aspects of planar Abelian electrodynamics and Condensed 
Matter phenomena, namely the up-to-date topic of high-Tc 
superconductivity. Still in this line, the study of the interaction 
between usual particles and Dirac-like objects may be useful in 
connection to low-energy problems.

\section{Acknowledgments}

E.M.C.A. is financially supported by Funda\c{c}\~ao de Amparo \`a Pesquisa
do Estado de S\~ao Paulo (FAPESP). W.A.M.M. is financially supported by 
Funda\c{c}\~ao de Amparo \`a Pesquisa do Estado de Minas Gerais (FAPEMIG). 
He also thanks CBPF and GFT/UCP where part of this work was done. This 
work was also partially supported by Conselho Nacional de Desenvolvimento 
Cient\'{\i}fico e Tecnol\'ogico (CNPq).\\

 \end{document}